# Inter-Tier Process Variation-Aware Monolithic 3D NoC Architectures

Shouvik Musavvir, *Student Member*, *IEEE*, Anwesha Chatterjee, *Student Member*, *IEEE*, Ryan Gary Kim, *Member, IEEE,* Dae Hyun Kim, *Member, IEEE*, Partha Pratim Pande, *Senior Member, IEEE*

*Abstract*— Monolithic 3D (M3D) technology enables high density integration, performance, and energy-efficiency by sequentially stacking tiers on top of each other. M3D-based network-on-chip (NoC) architectures can exploit these benefits by adopting tier partitioning for intra-router stages. However, conventional fabrication methods are infeasible for M3D-enabled designs due to temperature related issues. This has necessitated lower temperature and temperature-resilient techniques for M3D fabrication, leading to inferior performance of transistors in the top tier and interconnects in the bottom tier. The resulting inter-tier process variation leads to performance degradation of M3D-enabled NoCs. In this work, we demonstrate that without considering inter-tier process variation, an M3D-enabled NoC architecture overestimates the energy-delay-product (EDP) on average by 50.8% for a set of SPLASH-2 and PARSEC benchmarks. As a countermeasure, we adopt a process variation aware design approach. The proposed design and optimization method distribute the intra-router stages and inter-router links among the tiers to mitigate the adverse effects of process variation. Experimental results show that the NoC architecture under consideration improves the EDP by 27.4% on average across all benchmarks compared to the process-oblivious design.

## I. Introduction

Three-dimensional (3D) integrated circuits (ICs) have been shown to enable the design of high-performance and energy-efficient systems [1]. In particular, the network-on-chip (NoC) can heavily benefit from 3D integration as the communication backbone of manycore systems. By taking advantage of a third dimension, 3D NoCs provide a scalable communication fabric with lower hop-count, lower energy, and higher performance compared to their 2D counterparts [2].

Modern 3D integration processes have widely adopted through-silicon-via (TSV) technology to connect planar dies together. However, there are several significant challenges to TSV-based 3D integration. First, TSVs require additional fabrication steps like creating landing pads, wafer-thinning, and bonding [3]. These fabrication and packaging related challenges lead to lower yield rates and higher production costs for TSV-based designs [4]. Secondly, TSVs require a minimum keep-out-zone (KOZ) to reduce stress and coupling noise, introducing additional area overheads while undermining achievable integration density [5]. Thirdly, even though TSV-based vertical links can improve communication in 3D NoCs, they may fail due to crosstalk and electromigration [6].

Recently, monolithic 3D (M3D) integration has been proposed as an alternative to TSV-based designs. In M3D designs, multiple tiers are processed sequentially on the same die [7] and monolithic inter-tier vias (MIVs) are used as vertical links instead of TSVs. The physical dimensions of MIVs (~50nm × 100nm) are several orders of magnitude smaller than TSVs (1-3μm × 10-30μm) and are comparable to standard copper vias [8]. Similarly, the contact dimensions of M3D are much smaller (~100nm [9]) while TSV-based systems require contacts of 2-5μm. This allows us to achieve nanoscale contact pitch using M3D and attain the true benefit of vertical system integration. By facilitating nanoscale pitch, M3D enables us to examine gate- and block-level partitioning in circuits [7]. As a result, M3D offers much higher integration density and large reductions in total wire length over TSV-based counterparts. In addition, the direct wafer bonding technique in M3D achieves higher yields and lower costs compared to TSV-based integration [7], [10].

Naturally, NoC architectures can exploit the benefits of gate-/block-level partitioning in M3D integration by fabricating routers that span multiple tiers. In a recent study, M3D-enabled NoCs are shown to achieve 28% better energy efficiency compared to its TSV-based counterpart [11]. However, the investigations in [11] do not consider any M3D fabrication-related challenges or the benefit of lower interconnect capacitance from the reduced wire lengths in M3D designs [12].

While M3D architectures offer significant design flexibility and better energy efficiency compared to TSV-based designs, there are technology- and fabrication-related challenges that need to be addressed. M3D's sequential integration requires: 1) a low-temperature top-tier annealing process to prevent degradation in bottom-tier transistors [13]; and 2) temperature-resilient tungsten interconnects in the bottom tier to withstand high top-tier fabrication temperatures [7]. Unfortunately, these requirements degrade the transistors in the top tier and the interconnects in the bottom tier. Without considering these process-related effects it is likely that performance and energy efficiency will be overestimated at design time.

In this paper, we demonstrate, for the first time, the importance of including these two M3D design requirements (tungsten interconnects in the bottom tier and a low-

The manuscript was submitted on June 10, 2019. This work was supported in part by the US National Science Foundation (NSF) grants CNS-1564014, CCF 1514269, and USA Army Research Office grant W911NF-17-1-0485.

S. Musavvir, A. Chatterjee, D.H. Kim, and P.P. Pande are with the School of Electrical Engineering & Computer Science, Washington State University, Pullman, WA 99163 USA (e-mail: shouvik.musavvir@wsu.edu; anwesha.chatterjee@wsu.edu; daehyun.kim@wsu.edu; pande@wsu.edu).

R.G. Kim is with the Department of Electrical and Computer Engineering, Colorado State University, Fort Collins, CO 80524 USA (e-mail: Ryan.G.Kim@colostate.edu).

temperature top-tier annealing process) and the resulting inter-tier performance variation on the design, optimization, and efficiency of an M3D-enabled NoC architecture. Our main contributions of this paper are:

1) We demonstrate the necessity of including M3D process related effects in the design of NoCs. We show that a small-world NoC designed with M3D process parameters in mind lowers the energy-delay-product (EDP) with respect to the process-oblivious counterpart on average by 27.4% across all benchmarks under consideration.
2) We perform a detailed analysis of the effects of these M3D process related parameters and the benefits of partitioning intra-router stages across tiers on the design of process-aware NoCs. (*i.e.*, intra-router stage placement and inter-router link distribution).
3) We find that the distribution of intra-router pipelined stages and inter-router links among the M3D tiers is strongly dependent on the values of the process variation parameters. We also find that the distribution of the intra-router stages and inter-router links depends on the benchmarks under consideration. All these observations show and justify the need for the M3D process-aware 3D NoC design and optimization we propose in this paper.

The rest of the paper is organized as follows. The related work is discussed in Section II. Section III presents the challenges of M3D NoC design. In Section IV, we describe the problem setup and the proposed solution for the process variation induced performance variation. The experimental results are presented in Section V, and finally, we conclude the paper in Section VI.

## II. RELATED WORK

The merits of M3D-based designs have been explored in several works [14], [15]. M3D circuits provide reduced power, performance, and area compared to their 2D counterparts. Motivated by the promise of monolithic integration, the CELONCEL design framework was proposed to explore the advantage of transistor/gate-level partitioning and cell-on-cell stacking design for M3D integration [15]. It was found that the footprint and wirelength of M3D-based designs are reduced by 37.5% and 16.2% respectively over their planar counterparts at the 45 nm technology node. This results in a 6.2% reduction of overall delay for M3D designs for the same technology node. Moreover, performance improves for more advanced technology nodes such as 22 nm [16], 14 nm [8], and 7 nm [17]. The effects of the number of planar tiers, tier-level partitioning, and MIV insertion methodology on the performance of M3D-based ICs were analyzed [18]. As the number of MIVs in the design increases, the power saving improves as well. M3D systems enabled by nanotechnologies (N3XT) are proposed in [19]. N3XT employs recent nanotechnologies such as carbon nanotubes and M3D integration and achieves high-performance and energy efficiency.

Several works address the application of M3D technology to different types of circuits, *e.g.*, 3D FPGA [20], 3D DRAM [21], and 3D SRAM [22]. It is demonstrated that by using M3D technology, we can reduce the total area, path delay, and power consumption of the circuit. Similarly, the authors in [11] explored the design space of 3D NoCs and demonstrated the efficacy of M3D integration. However, all these studies considered ideal characteristics for the transistors and interconnects (*i.e.*, uniform delay and power consumption) across different tiers in the M3D ICs and ignored the effects of inter-tier process variation during design and evaluation.

The authors in [23] examined the effects of a realistic M3D fabrication process on the performance of M3D-enabled ICs. They found that the energy consumption and delay can increase significantly compared to an ideal M3D process. This study also evaluated the performance of M3D ICs that incorporate low-temperature annealing for top-tier transistors and tungsten interconnects in the bottom tier. Although M3D is a promising emerging interconnect technology, there is very little work on exploiting this to design a 3D NoC. The work in [11] incorporates router partitioning in the design of an M3D NoC but it does not consider the performance benefit of multitier router stages or inter-tier performance variation between tiers. Therefore, our aim is to create a design process that integrates inter-tier process variation and how the inclusion of inter-tier process variation (and lack thereof) impacts the design and optimization of an M3D NoC.

## III. CHALLENGES OF M3D-ENABLED DESIGNS

Although M3D enables higher integration density and better performance than TSV-based designs, fabrication challenges need to be addressed. For example, fabricating upper tiers becomes a major challenge due to the sequential tier synthesis in M3D integration and very thin inter-tier dielectric between the tiers. If ion implantation and annealing during top-tier fabrication uses the standard thermal budget (≈1050°C [7]), the high temperature can damage the underlying interconnects and transistors in the bottom tier. As a result, two techniques have been proposed to realize top-tier transistors without damaging lower tiers: solid phase epitaxy regrowth (SPER) [24] and laser annealing [25].

In [7], it has been shown that temperature must be kept below 650°C to prevent any damage to the lower-tier transistors. This is accomplished by both SPER, the ion implantation step is done at temperatures as low as 600°C [24]; and laser annealing, upper-tier transistors can be fabricated while only heating up the bottom tier to 500°C [25]. However, both procedures have disadvantages. Transistors created using SPER have three times higher source-drain resistance compared to conventional transistors [24]. Similarly, transistors fabricated using laser-based annealing have 16-28% lower on-current [13]. As a result, both processes introduce performance degradation for the top-tier transistors.

Unfortunately, although these temperatures are okay for lower-tier transistors, both SPER and laser annealing do not reduce the temperature enough to utilize copper back-end-of-line (BEOL) interconnects (the temperature must be kept within 400°C [7]). Therefore, a metal that can withstand higher temperatures such as tungsten is needed for BEOL interconnects in the bottom tier. However, tungsten has a higher resistivity than copper, which leads to inferior performance of bottom-tier interconnects [23].

These effects, degraded transistors in the upper tiers and higher resistivity interconnects in the lower tiers, can affect the design of NoCs. In particular, these inter-tier process variations cause non-uniform performance and energy consumption across the tiers. If these effects are not considered during design time, we may obtain overly optimistic latency and energy estimates and more importantly, sub-optimal NoC configurations. This motivates our work into formulating a new M3D-variation-aware NoC design problem and optimization.

## IV. Problem Formulation and Optimization

In a manycore architecture, our goal is to place cores, routers, and links efficiently and design the intra-router stages and inter-router links optimally to achieve the best NoC performance. We begin by discussing the optimization goals for a standard 2D NoC. Then, we extend this discussion to include M3D process variations and design considerations. Fig. 1 shows the overall flow of process variation-aware design of M3D NoCs. Finally, we present a framework that utilizes these optimization goals to design realistic M3D NoC architectures.

### A. Latency and Energy Modeling of NoCs

In this section, we define general models for NoC latency and energy that can be applied to both 2D and M3D systems. Later, we will provide specific details for 2D (Section IV.B) and M3D (Section IV.C) systems. We first model traffic-weighted NoC latency for a system with $N$ cores and $N$ routers as follows:

$$Latency = \sum_{i=1}^{N}\sum_{j=1}^{N}\sum_{r,u \in p(i,j)}\left(\sum_{m=1}^{S} t_{r,m} + l_u t_u\right) f_{i,j} \quad (1)$$

$p(i,j)$ gives the path between cores $i$ and $j$ (routers and links), where $r$ and $u$ are a router and a link along that path, respectively. The parameter $t_{r,m}$ is the intra-router stage delay for the $m^{th}$ router stage of router $r$ with $S$ router stages. It is important to note that $t_{r,m}$ depends on the number of ports and virtual channels associated with the particular router. The parameters $l_u$ and $t_u$ are the length of the interconnect and unit length delay of the inter-router link $u$, respectively. Lastly, $f_{i,j}$ is the frequency of interaction between cores $i$ and $j$. The latency captures the weighted sum of communication cost between every pair of cores.

Similarly, we model traffic-weighted NoC energy as follows:

$$Energy = \sum_{i=1}^{N}\sum_{j=1}^{N}\sum_{r,u \in p(i,j)}\left(\sum_{m=1}^{S} e_{r,m} + l_u e_u\right) f_{i,j} \quad (2)$$

Here, for the network path between any two cores $i$ and $j$, $e_{r,m}$ is the intra-router stage energy for the $m^{th}$ router stage of router $r$ with $S$ router stages. The parameter $e_u$ is the unit length energy of the inter-router link $u$.

In this work, our goal is to design energy-efficient and high-performance NoCs by minimizing NoC latency and energy simultaneously. Therefore, we need a unified optimization metric for NoC latency and energy. To consider the effect of both network latency and energy together, we use EDP as the relevant performance metric for NoC optimization. Using Eqs. (1) and (2), we can represent the EDP for NoC designs as follows:

$$EDP = Energy \cdot Latency \quad (3)$$

### B. 2D Router Delay and Energy Models

To determine the effects of M3D integration on the performance of an NoC, we first need to adopt an appropriate router model. In this work, we follow the virtual-channel router model proposed in [26]. However, it should be noted that any other router model can be adopted for this analysis (Eq. 1 and 2 are general in the number of stages). The virtual channel router has three pipelined stages, viz., virtual channel allocator, switch allocator, and crossbar traversal [26].

The delay of each stage depends on the number of router ports ($p$), flit size ($w$), and the number of virtual channels ($v$). All these parameters in turn depend on the adopted NoC architecture. Their relationship is given in Table I [26]. For a regular NoC architecture like mesh, the delay of a particular pipelined stage will be the same in each router since each router has the same number of ports (except the routers at the edge). However, small-world networks have already been shown to achieve significantly lower latency and energy consumption over mesh-based NoCs [11]. For these irregular small-world architectures, each router may have different number of ports. Hence, delay of each pipelined stage varies depending on the router configuration. Using the model given in Table I, we determine the delay ($t_{r,m}$) of each pipelined stage ($m$) in every router ($r$) in terms of fanout-of-four (FO4) delay.

TABLE I
PARAMETERIZED DELAY EQUATIONS FOR INTRA-ROUTER STAGES [26]

| Intra-router stage (m) | Delay (FO4) |
| --- | --- |
| Virtual channel allocator (1) | $33log_4(pv)+125/6$ |
| Switch allocator (2) | $28log_4p+35/2$ |
| Crossbar traversal (3) | $9log_8(w\lfloor p/2 \rfloor)+6\lfloor log_2p \rfloor+6$ |

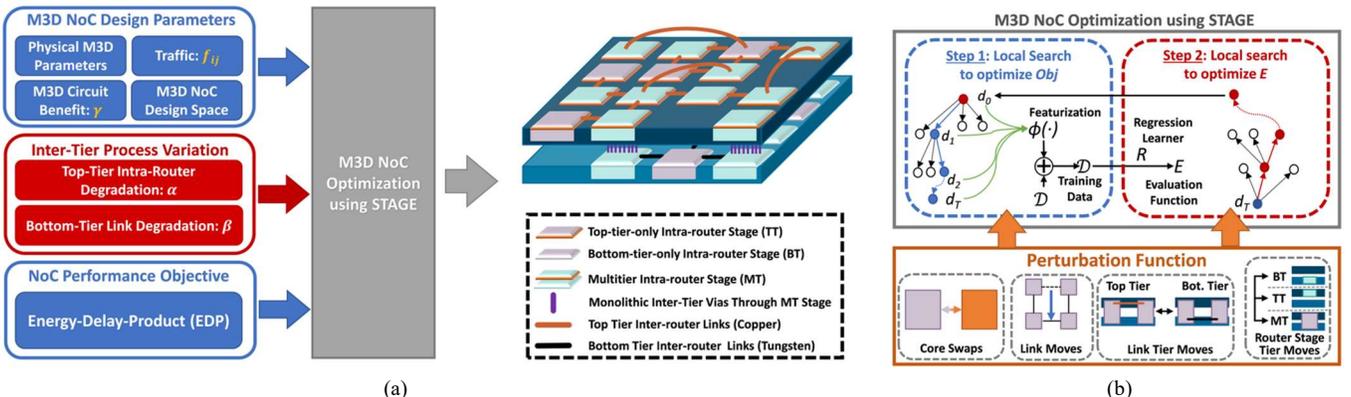

Fig. 1. (a) Illustration of the process variation-aware design methodology and an example small-world network-enabled 16-node M3D NoC architecture. The legend indicates different components of the NoC. (b) Illustration of the implementation of the STAGE algorithm for M3D NoC optimization.

Energy consumption of the routers ($e_{r,m}$) depends on the capacitance of the logic cells and the interconnect of each stage [27]. For ease of notation, we will denote $t_{r,m}$ and $e_{r,m}$ as $t_{2D-r,m}$ and $e_{2D-r,m}$, respectively for 2D planar designs. Also, since 2D systems use copper interconnects, $t_u = t_{Cu}$ and $e_u = e_{Cu}$ are the unit length delay and energy of transferring data through standard copper links, respectively. Using these models, we examine the effects of M3D integration on NoC design.

*C. Process Variation-Aware M3D NoC Design*

Although M3D integration allows us to build multitier routers, process variation from the M3D fabrication process causes the NoC to suffer from latency and energy penalties. As discussed in Section III, any router logic components in the top tier and inter-router links in the bottom tier will experience slowdowns due to process-related transistor degradation and the higher resistivity of tungsten, respectively. These effects could dominate the benefits obtained from tier partitioning, which would reduce the overall performance gain compared to its 2D counterpart. Therefore, in addition to placing the links between routers, it is our objective to choose the tiers for each router stage and link of the M3D-based NoC to minimize the EDP. Hence, we consider the following design choices separately for each router stage and link (Fig. 1(a)). Please note that this represents a high-level discussion, methods for determining delay, capacitance, etc. will be discussed later in the results section (Section V.C).

1) *Bottom-tier-only intra-router stage (BT)*: The performance of the router logic in BT is the same as the 2D counterpart because there is no degradation in the bottom-tier transistors. Therefore, $t_{r,m} = t_{2D-r,m}$ and $e_{r,m} = e_{2D-r,m}$.
2) *Top-tier-only intra-router stage (TT)*: As the transistors in the top tier are degraded, the FO4 delay of the router logic in TT will be larger than that in BT. To determine the delay of the router stage in TT, we need to determine the FO4 delay in TT in presence of transistor degradation. Hence the parameters $t_{r,m}$ and $e_{r,m}$ will vary depending on the degradation of the transistor on-current ($\alpha$). The intra-router delay for the router stage can be expressed as follows:

$$t_{r,m} = \frac{FO4_{TT}}{FO4_{2D}} \cdot t_{2D-r,m} \quad (4)$$

where $FO4_{TT}$ is the degraded FO4 delay in presence of $\alpha$ and $FO4_{2D}$ is the ideal FO4 delay for 2D design (Table I).

Transistor degradation will also increase the logic capacitance in TT stage [28]. As energy is proportional to the capacitance in the router stage, the stage energy is expressed as follows:

$$e_{r,m} = \frac{C_{TT,m}}{C_{2D,m}} \cdot e_{2D-r,m} \quad (5)$$

where $C_{TT,m}$ and $C_{2D,m}$ are the total capacitance of TT and 2D stage, respectively. $C_{TT,m}$ comprises of $C_{2D,m}$ and the incremental logic capacitance due to $\alpha$. It should be noted that the interconnect capacitance of the intra-router stage remains the same as the 2D counterpart.

3) *Multitier intra-router stage (MT):* By using block- and gate-level M3D integration, the interconnect length can be reduced to a factor of $1/\sqrt{T}$ for $T$-tier systems [12]. Therefore, although the change of logic capacitance is insignificant, the interconnect capacitance decreases by approximately $1 - 1/\sqrt{T}$ [12]. This results in an improvement of FO4 delay (denoted as $\gamma$) for a multitier design compared to the single-tier counterpart. In this work, for simplicity, we assume that the MT stages are equally distributed among the bottom and top tiers [29]. Hence, the delay and energy for the router stages are:

$$t_{r,m} = \gamma \cdot \left(\frac{1}{2} \cdot t_{2D-r,m} + \frac{1}{2}\frac{FO4_{TT}}{FO4_{2D}} \cdot t_{2D-r,m}\right) \quad (6)$$

$$e_{r,m} = \frac{C_{MT,m}}{C_{2D,m}} \cdot e_{2D-r,m} \quad (7)$$

Here, $C_{MT,m}$ is the total capacitance of the MT stage. The capacitance of the top tier logic of MT stage will increase as mentioned in TT whereas the interconnect capacitance will decrease compared to the 2D counterpart.

4) *Inter-Router Link Placement:* The delay and energy incurred by the inter-router links will depend on its tier placement. The inter-router links in the top tier use copper and do not suffer from any performance degradation (*i.e.*, $t_u = t_{Cu}$ and $e_u = e_{Cu}$). On the other hand, the bottom-tier tungsten interconnects exhibit higher resistance compared to the copper-based counterpart. Hence, the inter-router links in the bottom tier suffer from higher delay and energy consumption:

$$\begin{aligned} t_u &= t_W \\ e_u &= e_W \end{aligned} \quad (8)$$

where $t_W$ and $e_W$ are the delay and energy of tungsten links per unit length, respectively. Since tungsten has higher resistivity than copper, $t_W > t_{Cu}$ and $e_W > e_{Cu}$. We define the interconnect slowdown factor for tungsten interconnects as $\beta$, where $\beta$ is the degradation in the propagation delay of the tungsten interconnects compared to its copper counterpart. As the resistivity of nanoscale tungsten wires changes based on the dimensions and geometry [23], $\beta$ will vary too, which will be captured in Eqs. (1) and (2) by $t_W$ and $e_W$, respectively. Since these inter-router links are connected to the input and output stages of routers, a link and its respective router stages must be on the same tier. Therefore, during optimization (see Section IV.D), we constrain that top-tier links must be connected to TT or MT router stages and bottom-tier links must be connected to BT or MT router stages only.

*D. M3D NoC Design Optimization*

Since each router stage can be on different tiers (TT, BT, or MT) with either top- or bottom-tier links, the design space complexity increases dramatically from a 2D or TSV-based NoC to its M3D counterpart.

Such large search spaces make it difficult to utilize traditional optimization methods which rely on stochastic local exploration to reach the minima, *e.g.*, simulated annealing (SA) or genetic algorithms (GA). Therefore, intelligent search methods are necessary to reduce the run time and enhance scalability. We apply a machine learning based search technique, STAGE [30], to guide the search process. Prior works have already shown the efficacy of STAGE over SA and GA for different NoC architecture optimizations [11], [31].

STAGE works by utilizing past search trajectories to find better starting points. To accomplish this, STAGE iterates over two steps: (1) Hill climbing (or some other local search) to optimize $Obj$, the primary design objective; and (2) Hill climbing to optimize $E$, a learned evaluation function that predicts how promising a design is as a starting point for Step 1. We show these steps in Fig. 1(b).

1) *STAGE Step 1:* Similar to simple hill climbing or SA, the first step attempts to minimize the target function $Obj$ by making small steps (using a perturbation function $S$), accepting new designs if it reduces $Obj$, i.e., simple hill climbing. STAGE keeps track of all accepted designs in this search trajectory $(d_0, d_1, ..., d_T)$ and adds each design to a dataset $\mathcal{D}$, as a pair $(\phi(d), Obj(d_T))$. Here, $\phi(\cdot)$ is a function that extracts pertinent features from the design.
2) *STAGE Step 2:* Using a regression learner, $R$, Step 2 learns the evaluation function $E(\phi(d)) = R(\mathcal{D})$. This evaluation function tries to predict the EDP of the final design of Step 1 starting from a particular design $d$. Ultimately, $E$ can be used to predict the next best starting point for the search. Starting from the final design in Step 1, we use simple hill climbing to minimize $E$. This design is provided as the starting point to Step 1.
3) *STAGE Iteration:* STAGE iterates over Step 1 and Step 2 until the maximum number of iterations allowed ($Iter_{max}$) has been reached. After each iteration, we accumulate more data points in $\mathcal{D}$ and learn a more accurate $E$ which results in the search finding better designs. Our final output is the best design $d^*$ with minimum EDP.

In this work, we consider two different types of M3D NoCs, a process variation oblivious M3D NoC that uses MT for all router stages and uniformly distributes the links among layers [29] and our proposed process variation aware M3D NoC. To utilize STAGE for M3D NoC design, we use the following definitions. For the perturbation function, $S$, we consider two types of perturbations to move to neighboring designs in all M3D architectures: (i) swapping the position of two cores; and (ii) moving a link between a pair of routers with another of same length between two other routers. For process-aware M3D designs, we use two additional perturbations where: (iii) a router stage is switched among the three different stage types (TT, BT, and MT); or (iv) a link is switched between the top and bottom tier. For each call of the perturbation function S, the function chooses one among the available perturbations described above. We show these perturbation choices in Fig. 1(b).

The feature selection ($\phi(\cdot)$) is an important aspect of STAGE as relevant features allow us to learn an accurate evaluation function $E$. We use random forest regression to learn $E$. Since the EDP depends directly on the average network hops and traffic-weighted hop-count, we take account of these features. In addition, the clustering coefficient ($C_c$) is a measure of a router's connectivity with neighbors that can indicate better local communication. Hence, we consider these features, *i.e.*, average network hops, traffic-weighted hop-count, and the clustering coefficient. For process-aware M3D NoC designs we use two additional features: bottom-tier inter-router and top-tier intra-router performance penalty to account for the process variation effect on design.

## V. RESULTS AND ANALYSIS

In this section, we describe the experimental set up followed by the design considerations and architectural adjustments for our process-aware architecture. Then, we present a detailed EDP analysis of the M3D NoCs with process variation.

### A. Experimental Setup

In this work, we consider a 64-core system where each core is associated with a dedicated router. We use GEM5 [32], a full system simulator, to obtain detailed processor- and network-level information. Using Gem5's full-system mode, we simulate x86 cores running Linux. We use the MOESI_CMP_directory cache coherence protocol. Each core consists of private 64KB L1 instruction and data caches and a shared 8MB L2 cache. We consider four SPLASH-2 benchmarks (FFT, RADIX, LU, and WATER) [33] and four PARSEC benchmarks (DEDUP, VIPS, CANNEAL (CAN), and FLUIDANIMATE (FLUID)) [34]. These benchmarks are selected because they vary widely in communication and computation patterns.

### B. NoC Design and Baseline

For each router, we use the three-stage model shown in Table I [26]. Each router has four virtual channels ($v$=4) per port. Each packet contains six flits and each flit consists of 32 bits ($w$=32). These routers are synthesized from an RTL-level design using a TSMC 28 nm CMOS process in Synopsys$^{TM}$ Design Vision. The MIV physical dimensions are 50 nm × 50 nm × 44 nm (diameter × depth × pitch) [8]. It should be noted that as the dimensions of MIV are comparable to that of local via, using standard 2D cells in Synopsys to synthesize the NoC routers does not add any additional overhead [8]. Energy consumption of the NoC links is determined using HSPICE simulations, taking their lengths and resistivity (the bottom tier uses tungsten while the top tier uses copper) into consideration.

For the NoC topology, we consider two cases: small-world network-enabled 3D NoC (SWNoC) and traditional 3D mesh. We create the SWNoC architecture by following a power-law-based link distribution as elaborated in [31]. It should be noted that we have already demonstrated in our prior works that SWNoC outperforms any other regular and application-specific 3D NoC architectures [11], [31]. For 3D mesh, we use standard XYZ-dimension-order routing. Since small-world networks are irregular topologies, we adopt the topology-agnostic adaptive layered shortest path routing (ALASH) for SWNoCs [35].

### C. M3D Transistor and Interconnect Characteristics

In this work, we consider a two-tier M3D process. In order to deliver a thorough analysis of M3D NoCs, we must first provide a practical range for the degradation of transistor on-current ($\alpha$), slowdown factor due to tungsten interconnects ($\beta$), and M3D improvement in FO4 delay ($\gamma$) parameters (see Section IV.C) and determine their effects on the design and optimization of M3D NoCs.

For $\alpha$, prior work has reported a maximum of 20% degradation in the top-tier transistors [36]. Hence, we can consider $\alpha = [0.05, 0.20]$ in 0.05 increments. The parameter $\alpha$ affects $FO4_{TT}$, $C_{TT,m}$, and $C_{MT,m}$, and hence, the energy and delay of TT and MT intra-router stages. We used Cadence

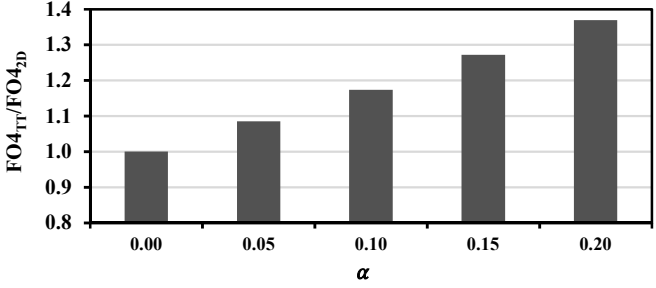

Fig. 2. $FO4_{TT}/FO4_{2D}$ for different values of $\alpha$.

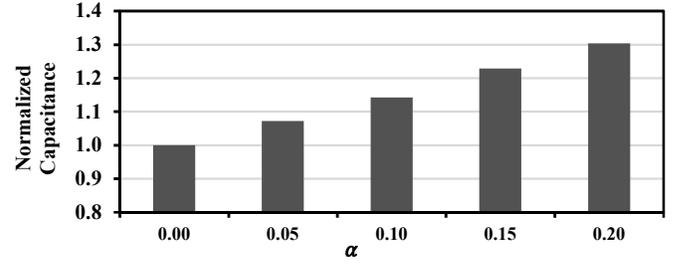

Fig. 3. Normalized transistor capacitance for different values of $\alpha$. Capacitance is normalized with respect to the nominal transistor.

Virtuoso to determine $FO4_{TT}$ in presence of $\alpha$. Fig. 2 shows the ratio of $FO4_{TT}$ to $FO4_{2D}$ for different values of $\alpha$. Here, $\alpha = 0$ is the case for nominal 2D transistors. For every 5% increment in $\alpha$, the FO4 delay degrades by approximately 9%. The increment of top-tier logic capacitance for TT ($C_{TT,m}$ in Eq. (5)) and MT ($C_{MT,m}$ in Eq. (7)) is estimated from the relationship between transistor on-current and capacitance presented in [28]. In Fig. 3, the normalized capacitance for transistors is plotted for different values of $\alpha$. Here, the capacitance is normalized with respect to a nominal 2D transistor ($\alpha = 0$). As expected, the capacitance increases almost linearly with f $\alpha$.

For $\beta$, the impact of tungsten interconnects in the bottom tier depends on the metal layer and technology node [23]. We consider tungsten interconnects in metal layers 3 through 10 for a TSMC 28nm process. For each layer we characterize the delay of the tungsten interconnect by changing the resistivity [23] in Cadence Virtuoso. Our experimental analysis shows that tungsten interconnects introduce 10-30% additional propagation delay per unit length. Hence, we use $\beta = [0.10, 0.30]$ in 0.10 increments.

For $\gamma$, M3D designs achieve up to 25% improvement in clock frequency compared to their 2D counterpart [37]. Thus, we use two values for $\gamma$ (0.10 and 0.20) in our work. To determine the MT stage energy in Eq. (7), we find $C_{MT,m}$, which must consider the increased logic capacitance at the top tier due to $\alpha$ [28] and the lower interconnect capacitance due to tier partitioning.

### D. Router Stage and Link Distribution with Process Variation

Under ideal conditions ($\alpha=0$, $\beta=0$), naturally, all intra-router stages in an M3D NoC will be multitier and the links will be placed evenly in both tiers. Unfortunately, as discussed in Section IV.C, top-tier transistor degradation ($\alpha > 0$) and bottom-tier interconnect degradation ($\beta > 0$) cause significant slowdowns in the M3D NoC. Moreover, these slowdowns are not uniform across different tiers. In addition to these non-uniform effects of M3D process variation, the delay and capacitance of each intra-router stage varies with the number of bits per flit, ports, and VCs in the router (see Table I). Thus, to minimize the overall EDP for a given system configuration (number of bits per flit, number of ports, number of VCs, packet size, $\alpha$, $\beta$, and $\gamma$), we should optimize the distribution of the intra-router stages and inter-router links. We present the analysis for SWNoCs followed by mesh based NoCs.

#### 1) M3D SWNoC Architecture Optimization

In SWNoCs, the distribution of intra-router stages and links depends on the process variation effects and traffic distribution. Across a range of $\alpha$, $\beta$, and $\gamma$, Fig. 4 shows the tier-wise distribution of all intra-router stages optimized for the CAN benchmark as an example. Since there is a trade-off between BT and MT in terms of link versus logic degradation, we see different distributions of BT and MT. Although the logic in the top tier of an MT stage has longer delay and larger capacitance than the nominal values, the overall wirelength is shorter than sole BT stage. Therefore, some of the router stages can benefit from tier partitioning depending on $\alpha, \beta$, and $\gamma$. For example, Fig. 4 shows that approximately 80% of the router stages are MT and 20% are BT when $(\alpha, \beta, \gamma)$ is (0.05, 0.1, 0.1). However, if $\alpha$ increases, more stages are chosen to be BT to avoid the intra-router performance penalty originating from the top-tier transistor degradation as explained in Eqs. (4) and (5). On the other hand, since $\gamma$ improves the MT intra-router stage delay (Eq. (6)), there are more MT stages at higher values of $\gamma$. Notably, the result shows that the SWNoCs do not have any TT stages. In MT, only half of the logic cells suffer from transistor degradation. Whereas all logic cells in TT experience transistor slowdown. Moreover, the speedup due to multitier logic ($\gamma$) results in superior performance of MT compared to TT. As a result, the optimization always chooses MT over TT for all intra-router stages.

It should be noted that the distribution of different types of intra-router stages depends on their circuit composition. Since the crossbar stage is dominated by the interconnect capacitance [27] and interconnects are heavily reduced in MT (interconnect capacitance decreases by 29.3% for the two-tier system), the energy saving in the interconnects offsets the transistor slowdown. In Fig. 5, it can be seen that besides a few crossbar stages at low values of $\beta$, all crossbar stages are MT for every combination of $\alpha$, $\beta$, and $\gamma$.

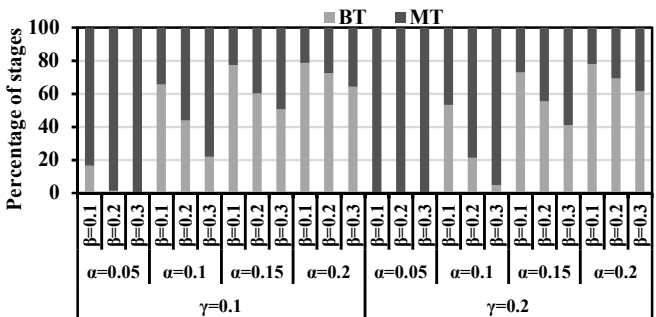

Fig. 4. Tier distribution of all intra-router stages for SWNoC (CAN).

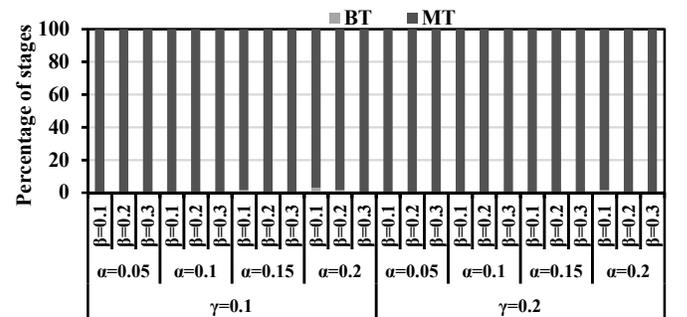

Fig. 5. Tier distribution of the crossbar stages for SWNoC (CAN).

As mentioned in Section IV.C, there is no effect of $\beta$ on the placement of the crossbar stage (it is not connected directly to an inter-router link). However, the switch allocator (SWA) and virtual channel allocator (VCA) stages are connected to the router ports, which in turn are connected to the inter-router links. As discussed in Section IV.C, the $\alpha$ and $\gamma$ parameters affect the intra-router stages and $\beta$ slows down the inter-router links in the bottom tier. Thus, all three parameters $\alpha$, $\beta$, and $\gamma$ influence the distribution of the SWA and VCA stages and the inter-router links associated with them. Figs. 6 and 7 show the distribution of the SWA and VCA stages, and inter-router links, respectively for different values of $\alpha$, $\beta$, and $\gamma$ for the CAN benchmark. As $\beta$ increases, the number of BT stages decreases (Fig. 6) and more links are placed at the top tier (Fig. 7) to avoid the interconnect performance penalty. For the links in the top tier, the associated stages must become MT or TT (TT is never chosen since MT is always better as discussed earlier). So, in Fig. 6, the number of MT stages increases with the rise of $\beta$. Conversely, as $\alpha$ increases, MT stages experience more delay and energy degradation. Hence, more stages (Fig. 6) and their respective links (Fig. 7) are placed at the bottom tier to avoid the transistor degradation. For $\gamma = 0.1$, 95.9% of the stages and 97.8% of the links are placed in the bottom tier when we consider the highest value of $\alpha$ ($\alpha = 0.2$) and the lowest value of $\beta$ ($\beta = 0.1$). Alternatively, all the intra-router stages are MT and all the links are placed in the top tier for the lowest value of $\alpha$ ($\alpha = 0.05$) and the highest value of $\beta$ ($\beta = 0.3$). Overall, the router stages and the inter-router links are distributed to reduce the effects of $\alpha$ and $\beta$ as much as possible. We can also observe the effect of $\gamma$ in Figs. 6 and 7. On average (considering different $\alpha$ and $\beta$), the number of MT stages and top-tier links increase by 9.6% and 8%, respectively when $\gamma$ increases from 0.1 to 0.2. Hence, considering the effects of various process variation parameters, the NoC router stages can be placed on a single tier (BT) or can be distributed over multiple tiers (MT) to optimize EDP.

In Fig. 8, the tier-wise distribution for the stages connected to a link of particular length is plotted for the CAN benchmark as an example. The link length is expressed in terms of Manhattan distance. As we can see, the placement of SWA and VCA stages is associated with the length of the links. The inter-router traversal penalty at the bottom tier is proportional to the link length (Eq. (8)). Hence, the long-range links are placed mostly in the top tier to avoid the performance penalty ($\beta$) while the shorter links along with its respective SWA and VCA stages are placed in the bottom tier. Here, the stages connected to longer links favor MT over BT.

We also found similar trends of the stage and link distribution in all the benchmarks. We plot the input/output stage (SWA and VCA) and link distribution of SWNoCs in Figs. 9 and 10 for representative values $\alpha = 0.1$ and $\gamma = 0.1$ with varying $\beta$. As previously discussed, there are zero TT stages and the number of BT stages and bottom-tier links decrease with increasing $\beta$ across all benchmarks. The placement of intra-router stages connected to links (SWA and VCA) and inter-router links varies depending on the traffic characteristics of the specific benchmark. To understand this benchmark dependence, we analyzed RADIX (high percentage of BT and bottom-tier links) and WATER (low percentage of BT and bottom-tier links), two benchmarks that exhibit distinct link distribution trends. In Fig. 11, we show the percentage of traffic exchanged between two routers separated by a particular Manhattan distance. For two routers separated by one Manhattan distance, the traffic exchanged is significantly more for RADIX compared to WATER (18.6% on average). Moreover, RADIX has almost no traffic between routers separated by a Manhattan distance greater than three. Since traffic in RADIX doesn't have to travel physically far, much of the RADIX traffic takes short-distance links. As we found in the analysis of Fig. 8, this influences the network to have short bottom-tier interconnects. On the other hand, WATER has more traffic that needs to travel further, causing less bottom-tier links, especially for higher values of $\beta$. In fact, all the intra-router stages are BT and all the inter-router links are placed at the bottom tier for RADIX ($\alpha = 0.1$, $\gamma = 0.1$ and $\beta = 0.1$) as 77.6% of the total traffic is exchanged between routers separated by one Manhattan distance (Fig. 11). Hence, the link and stage distribution depend on the degree of process variation and traffic pattern of the respective benchmark.

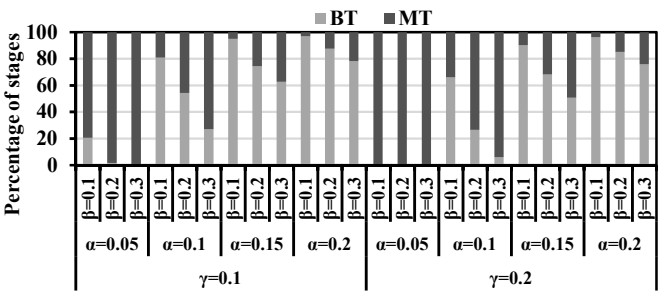

Fig. 6. Tier distribution of the SWA and VCA stages for SWNoC (CAN).

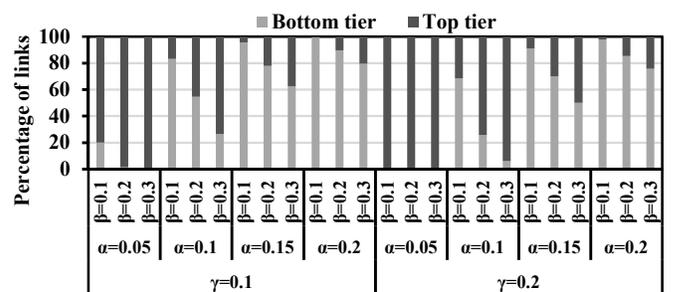

Fig. 7. Tier distribution of links for SWNoC (CAN).

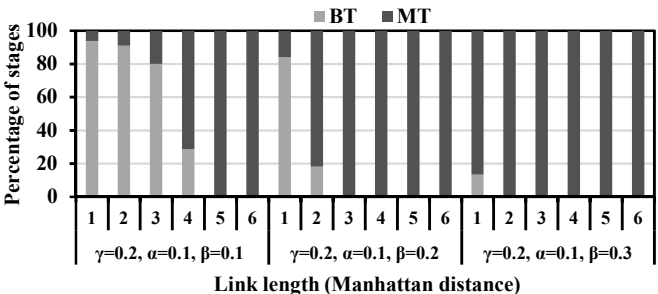

Fig. 8. Tier distribution of the SWA and VCA stages connected to a link of particular length for SWNoC (CAN).

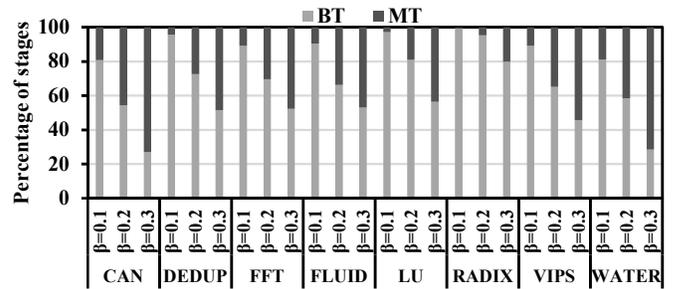

Fig. 9. Tier distribution of the SWA and VCA stages for SWNoC considering all benchmarks ($\alpha=0.1, \gamma=0.1$).

*2) Optimization of Mesh-based NoC*

So far, we have considered the SWNoC architecture to thoroughly study the effects of process variation on the NoC router configuration. To study the effects of the process variation on a regular NoC architecture, we undertake the same analysis on an equivalent mesh NoC of the same size. Although the crossbar stage distribution of mesh NoCs is similar to that of SWNoCs, the input/output stages' and links' tier placement are different. We show the distribution of stages (SWA and VCA) and links for a mesh NoC in Figs. 12 and 13, respectively considering the CAN benchmark. Compared to the stage (Fig. 6) and link (Fig. 7) distributions in SWNoC, the SWA and VCA stages, and links in mesh NoC favor the bottom tier. This is attributed to the fact that a mesh NoC consists of only short-range links between adjacent routers causing the router logic to have greater influence on delay and energy. Therefore, in order to avoid the top-tier transistor penalty (except for low values of $\alpha$), links are placed in the bottom tier. On the other hand, SWNoCs contain both short- and long-range links. The long-range links placed between non-adjacent routers incur more delay and energy overhead if they are placed in the bottom tier.

The general characteristics of process-aware design in SWNoCs are also present in their mesh counterparts. As $\gamma$ increases, the number of MT stages also increases as seen in Fig. 12. Similarly, the number of BT stages (Fig. 12) and bottom-tier links (Fig. 13) increase with lower values of $\beta$ or higher values of $\alpha$. Hence, our process-aware design accounts for the effects of $\alpha$ and $\beta$ for both SWNoC and mesh NoC.

*E. Process-Oblivious vs. Process-Aware M3D NoC*

In this section, our aim is to demonstrate the advantages of designing the M3D NoC when considering the effects of process variation. As explained above, when we consider the effects of process variation, the router stages and inter-router links need to be distributed suitably among the tiers. We call this M3D NoC the process-aware architecture (M3D-PA). On the contrary, if we do not consider the effects of process variation (by assuming $\alpha=0$, $\beta=0$) while designing the M3D NoC, every router would be multitier to take advantage of the performance benefit due to $\gamma$, we call this M3D NoC the process-oblivious M3D NoC (M3D-PO). Our aim is to quantify the benefits of our process-aware design compared to its process-oblivious counterpart.

Figs. 14 and 15 show the EDP of the SWNoC and mesh NoC, respectively for the CAN benchmark. The EDP is normalized with respect to the M3D-PO design under ideal conditions ($\alpha = 0$, $\beta = 0$). For SWNoCs at the lowest value of $\alpha$ ($\alpha = 0.05$), the M3D-PA design improves the EDP by 10.7% and 9.1% on average considering all values of $\beta$ over its M3D-PO counterpart for $\gamma = 0.1$ and $\gamma = 0.2$, respectively. Since the stage distribution of M3D-PO and M3D-PA design is similar for $\alpha=0.05$, the difference in EDP between these two design approaches is low. In addition, at low values of $\alpha$ most of the stages are MT (Fig. 4), this allows the optimization to utilize the top layer for links, reducing the effects of beta on EDP.

As the value of $\alpha$ increases, the MT router stages get increasingly penalized, so the M3D-PA designs use fewer MT and more BT stages as shown in Fig. 4. Therefore, there is more opportunity to make better decisions by establishing the tradeoff between $\alpha$ and $\beta$. However, since the M3D-PO designs do not consider the process variation, the EDP difference between the M3D-PA and M3D-PO designs becomes more prominent. For the most severe process variation ($\alpha=0.2$, $\beta=0.3$), the M3D-PA SWNoCs outperform the M3D-PO counterparts by 43.9% and 37.6% for $\gamma=0.1$ (Fig. 14(a)) and $\gamma=0.2$ (Fig. 14(b)), respectively.

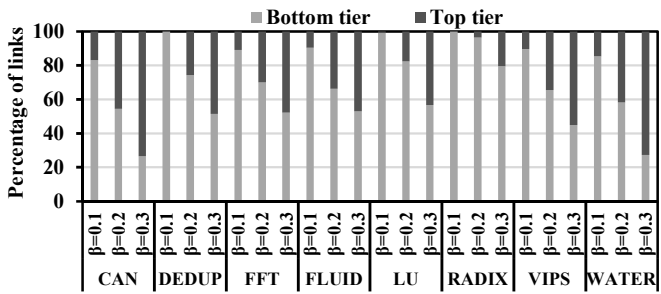

Fig. 10. Tier distribution of links for SWNoC considering all benchmarks ($\alpha=0.1$, $\gamma=0.1$).

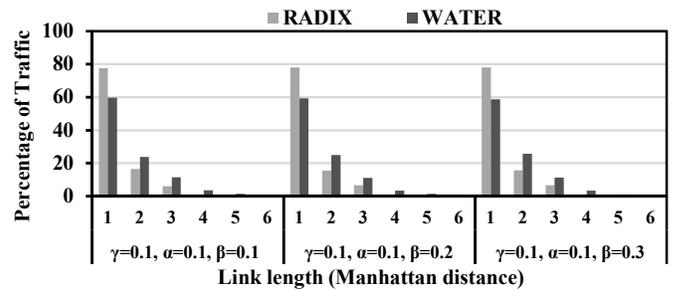

Fig. 11. Percentage of traffic exchanged between any two routers as a function of Manhattan distance for RADIX and WATER.

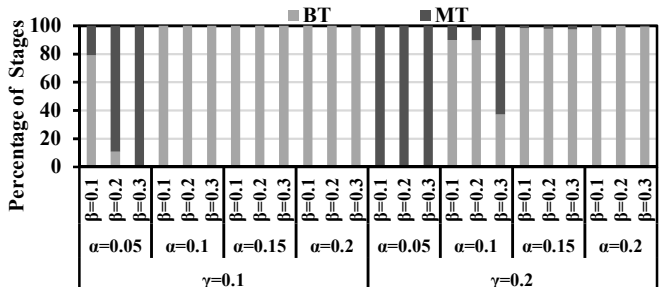

Fig. 12. Tier distribution of stages for mesh NoC (CAN).

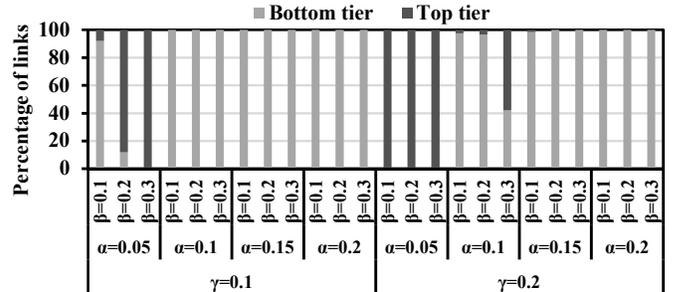

Fig. 13. Tier distribution of links for mesh NoC (CAN).

We observe similar trends in the EDP distribution of the mesh NoCs in Fig. 15. In fact, we save more EDP for the M3D-PA mesh NoCs with respect to the M3D-PO counterpart than the SWNoC. For the maximum process variation, the process-aware designs outperform the process-oblivious counterpart by 69.6% and 64.2%, respectively, for $\gamma=0.1$ (Fig. 15(a)) and $\gamma=0.2$ (Fig. 15(b)). In the mesh NoCs, we need more hops to traverse between a pair of source and destination routers on average. This results in passing through more intra-router stages and inter-router links which relates to more opportunities to make the appropriate tradeoffs due to the process variation parameters ($\alpha$ and $\beta$).

We show the EDP of all benchmarks in Fig. 16 for the SWNoC and the mesh NoC. We chose three different process variation combinations that cover the range of possible values: $\alpha=0.1, \beta=0.1$ (LOW), $\alpha=0.15, \beta=0.2$ (MED), and $\alpha=0.2, \beta=0.3$ (HIGH) to observe the effects of different levels of process variations while keeping $\gamma$ at 0.1. As discussed above, the EDP of the M3D-PO design deteriorates as the value of $\alpha$ and $\beta$ increases. For the SWNoCs (Fig. 16(a)), on average, the M3D-PA saves 19.6%, 33.1%, and 48.7% EDP compared to its M3D-PO counterpart for LOW, MED, and HIGH respectively. For the mesh NoCs (Fig. 16(b)), the M3D-PA design saves 27.5%, 47.9%, and 70.2% EDP on average compared to its M3D-PO counterpart for LOW, MED, and HIGH respectively.

*F. EDP Comparison Between TSV- and M3D-Based SWNoCs*

To complete the analysis, we compare the M3D-PA SWNoC with respect to the TSV-based counterpart of the same size [31]. In Fig. 17, we present the EDP of TSV and M3D-PA SWNoCs. The EDP is normalized with respect to the TSV-based design. Here we consider the maximum effect of process variation ($\alpha=0.2, \beta=0.3$) and the lowest value of $\gamma$ ($\gamma=0.1$). It is evident from the figure that even in the worst case for process variations, M3D-based designs still reduce the EDP by 11.7% on average compared to TSV based designs for all benchmarks.

VI. CONCLUSION

While M3D-integration offers high performance and energy efficiency, fabrication-related challenges pose major concerns to achieve desirable performance levels. The process-induced performance degradation of the transistors and interconnects introduce significant performance and energy overheads for M3D-enabled NoCs. Our analysis shows that the SWNoC designed without considering the process variation underestimates the EDP by at least 18.8% and at most 83.7% depending on the process variation parameters. Thus, process variation aware design is a must for realistic M3D NoC architectures. In this work, we incorporate both top-tier transistor slow-down and bottom-tier interconnect degradation in the NoC design process. Our proposed design reduces the performance degradation of the M3D NoC by suitably distributing the intra-router stages and inter-router links among the M3D tiers. The process-aware design improves the EDP of SWNoC by at least 7.2% and up to 48.7% compared to the process-oblivious design approach for the best- and worst-case of process variation, respectively.

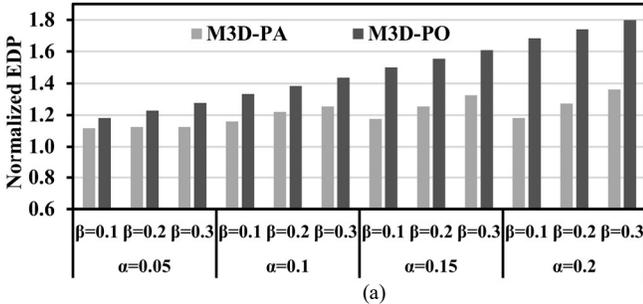 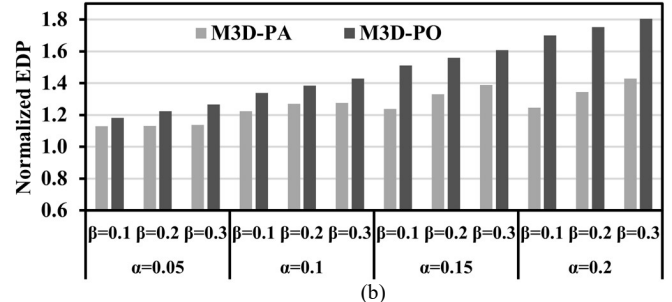

Fig. 14. EDP for SWNoCs for a) $\gamma=0.1$ and b) $\gamma=0.2$. (CAN). EDP is normalized with respect to that of process-oblivious design under ideal conditions ($\alpha=0, \beta=0$).

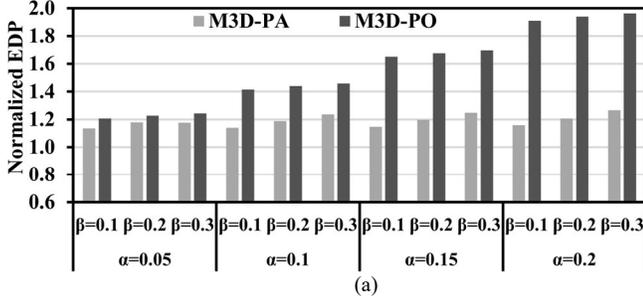 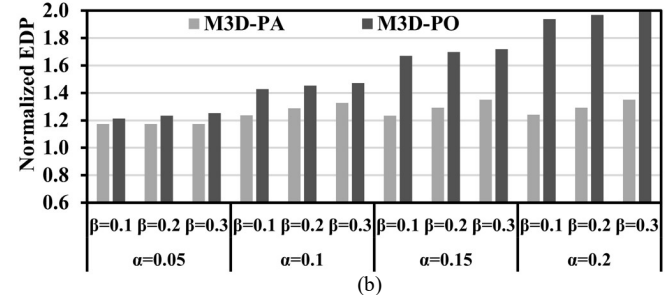

Fig. 15. EDP for mesh NoCs for a) $\gamma=0.1$ and b) $\gamma=0.2$. (CAN). EDP is normalized with respect to that of process-oblivious design under ideal conditions ($\alpha=0, \beta=0$).

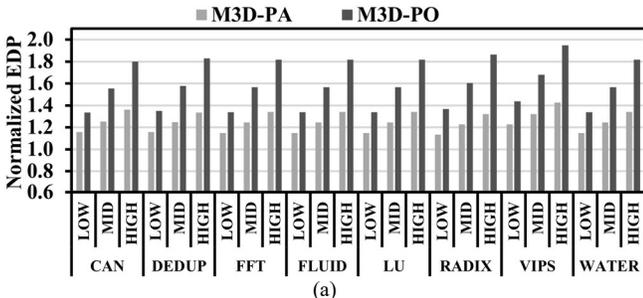 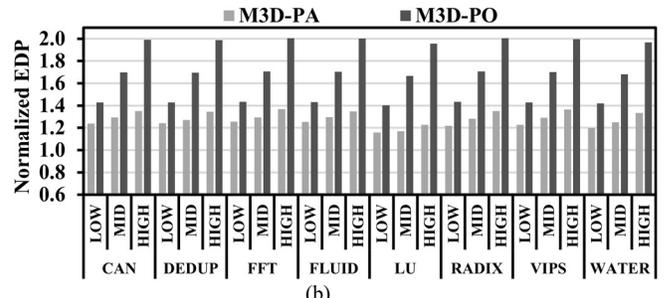

Fig. 16. EDP for a) SWNoCs and b) mesh NoCs considering all benchmarks ($\gamma=0.1$). EDP is normalized with respect to that of process-oblivious design under ideal conditions ($\alpha=0, \beta=0$).

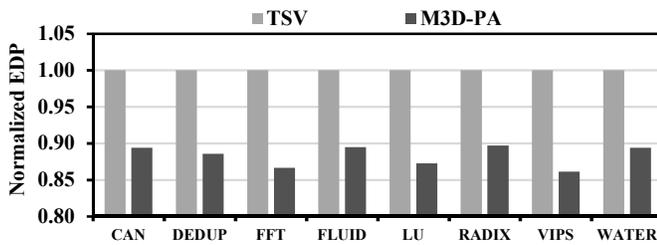

Fig. 17. EDP of TSV- and M3D-enabled ($\gamma=0.1, \alpha=0.2, \beta=0.3$) SWNoCs. EDP is normalized with respect to that of the TSV-based design.